\documentclass[conference]{IEEEtran}
\IEEEoverridecommandlockouts
\usepackage{cite}
\usepackage{amsmath,amssymb,amsfonts}
\usepackage{algorithmic}
\usepackage{graphicx}
\usepackage{textcomp}
\usepackage{xcolor}
\usepackage{caption}
\usepackage{graphicx}
\usepackage{float} 
\usepackage{subcaption}
\usepackage{pifont}
\usepackage[linesnumbered,ruled,vlined]{algorithm2e}
\usepackage{url}
\usepackage{multirow}
\def\BibTeX{{\rm B\kern-.05em{\sc i\kern-.025em b}\kern-.08em
    T\kern-.1667em\lower.7ex\hbox{E}\kern-.125emX}}
\begin{document}

\title{Enabling Efficient Batch Serving for LMaaS via Generation Length Prediction
}
\author{\IEEEauthorblockN{Ke Cheng\textsuperscript{1,}\IEEEauthorrefmark{1}, Wen Hu\textsuperscript{2}, Zhi Wang\textsuperscript{2}, Peng Du\textsuperscript{2}, Jianguo Li\textsuperscript{2,}\IEEEauthorrefmark{2}, Sheng Zhang\textsuperscript{1,}\IEEEauthorrefmark{2}}
\IEEEauthorblockA{\textsuperscript{1}State Key Laboratory for Novel Software Technology, Nanjing University, Nanjing, China\\
	\textsuperscript{2}Ant Group, Hangzhou, China \\
ketonmi@outlook.com, \{huwen.hu,wangchun.wz,dp333152,lijg.zero\}@antgroup.com, sheng@nju.edu.cn}
\thanks{\IEEEauthorrefmark{1} Work done during an internship at Ant Group.}
\thanks{\IEEEauthorrefmark{2} Jianguo Li and Sheng Zhang are corresponding authors.}
\vspace{-1cm} 
}

\maketitle

\begin{abstract}
Nowadays, large language models (LLMs) are published as a service and can be accessed by various applications via APIs, also known as language-model-as-a-service (LMaaS).
Without knowing the generation length of requests, existing serving systems serve requests in a first-come, first-served (FCFS) manner with a fixed batch size, which leads to two problems that affect batch serving efficiency. First, the generation lengths of requests in a batch vary, and requests with short generation lengths must wait for requests with long generation lengths to finish during the batch serving procedure. Second, requests with longer generation lengths consume more memory during serving. Without knowing the generation lengths of batched requests, the batch size is always set small to avoid the out-of-memory (OOM) error, thus preventing the GPU from being fully utilized.
In this paper, we find that a significant number of popular applications in the LMaaS scenario have a positive correlation between the generation length and the length of raw user input. Based on this observation, we propose Magnus, which can accurately predict the request generation length with the user input length, application-level, and user-level semantic features. Accordingly, Magnus can achieve high request throughput by batching requests of similar generation lengths together with adaptive batch sizes.  Besides, Magnus can also schedule batches with the highest response ratio next (HRRN) policy to reduce request response time. Experiments conducted on our testbed show that Magnus improves request throughput by up to 234\% and reduces response time by up to 89.7\% compared to baselines.
\end{abstract}


\section{Introduction}
\label{sec:introduction}
As the parameter scale increases, transformer-based generative large language models (LLMs) have shown strong power on a wide range of natural language processing (NLP) tasks \cite{gpt, opt, chatglm, llama, llama2, qwen, baichuan}, which benefit numerous applications. However, most application developers cannot afford to train and deploy LLMs due to the prohibitive cost. Therefore, AI giants such as OpenAI, Google, and Baidu release their LLMs as a service and allow developers to access their models through APIs, which is called language-model-as-a-service (LMaaS) \cite{sun2022black}.

\begin{figure}[t]
	\setlength{\belowcaptionskip}{-0.5cm} 
	\centering
	\includegraphics[width=0.9\linewidth]{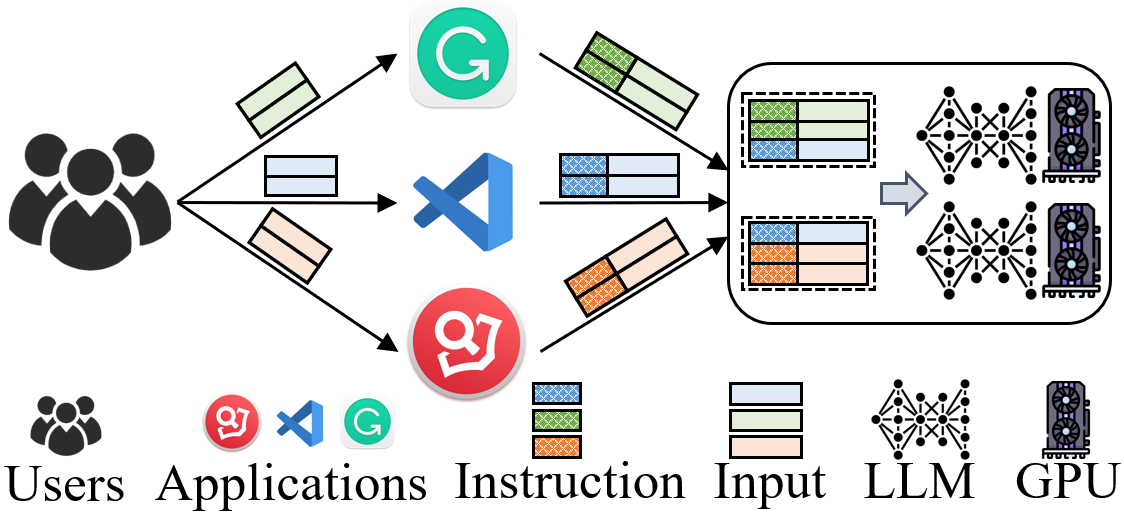}
	\caption{LMaaS scenario.}
	\label{fig:lmaas_scenario}
\end{figure}

Fig. \ref{fig:lmaas_scenario} illustrates the LMaaS scenario. On the application side, the users' input texts are attached with specific instructions as requests which are sent to the LLM service for serving. For example, the code assistant plugin on VSCode can implement its bug-fixing feature by prefixing the user's input code with the instruction "Fix bugs in the following code and output the fixed code:" as a request and feeding the request to the LLM service. 
On the LLM service side, the requests from different applications are mixed in batches and processed by LLM instances deployed on accelerators such as the graphics processing unit (GPU).

From the perspective of LLM service providers, it is desirable to reduce request response time and increase request throughput to improve quality of service (QoS) and serve more end users. However, without knowing the generation length of requests, existing serving systems such as Tensorflow Serving \cite{tensorflowserving} and Triton Inference Server \cite{triton} serve requests with a fixed batch size in a first-come first-served (FCFS) manner. This leads to two problems that hurt the serving efficiency. 
First, in a batch, requests with short generation lengths must wait for requests with long generation lengths to complete before they can be returned together. During waiting, the completed requests still participate in the computation and generate invalid tokens, leading to severe computational waste. Second, because requests with longer generation lengths generate more key-value cache which consumes more GPU memory, without knowing the generation length of the request, existing serving systems always use a small batch size to avoid out-of-memory (OOM) errors. Therefore, the parallel computing capability of GPUs cannot be fully exploited.

\begin{figure*}[t]
	\setlength{\belowcaptionskip}{-0.25cm} 
	\centering
	\begin{subfigure}{0.157\linewidth}
		\centering
		\includegraphics[width=1\linewidth]{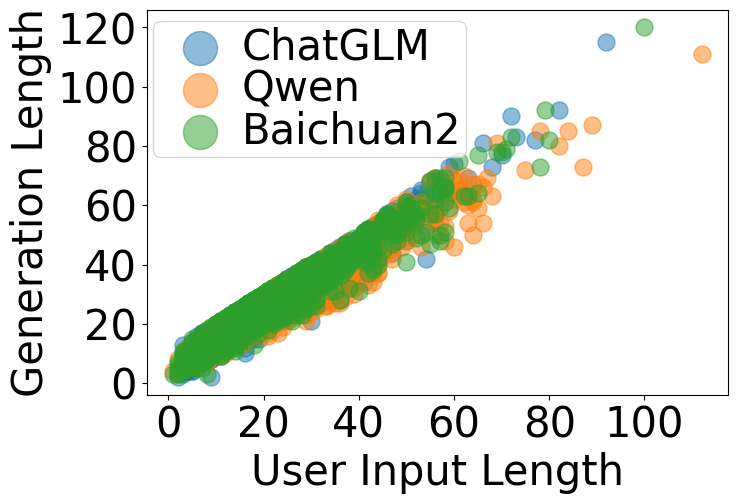}
		\caption{MT}
		\label{fig:corr_mt}
	\end{subfigure}
	\centering
	\begin{subfigure}{0.1615\linewidth}
		\centering
		\includegraphics[width=1\linewidth]{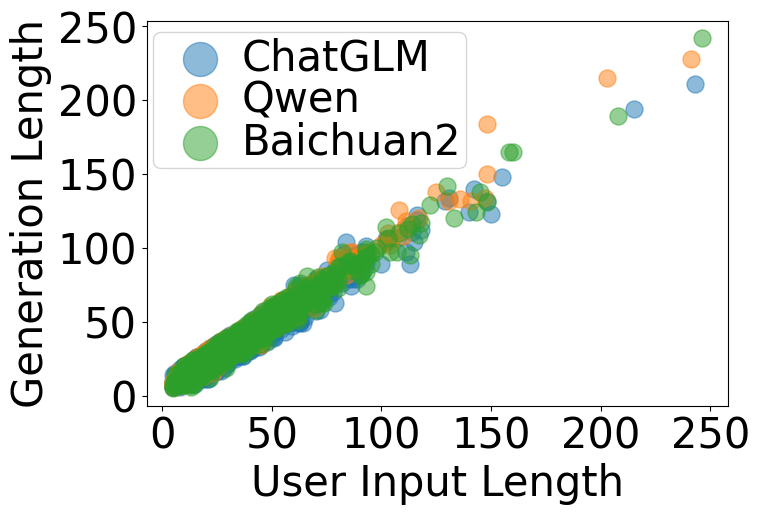}
		\caption{GC}
		\label{fig:corr_gc}
	\end{subfigure}
	\centering
	\begin{subfigure}{0.151\linewidth}
		\centering
		\includegraphics[width=1\linewidth]{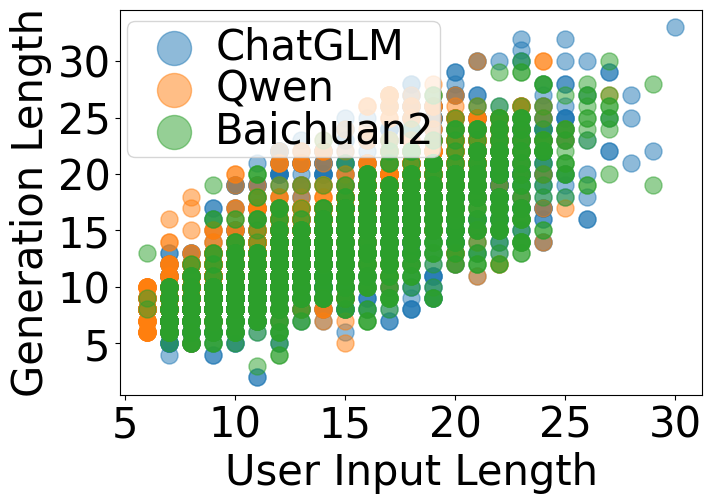}
		\caption{TD}
		\label{fig:corr_td}
	\end{subfigure}
	\centering
	\begin{subfigure}{0.1565\linewidth}
		\centering
		\includegraphics[width=1\linewidth]{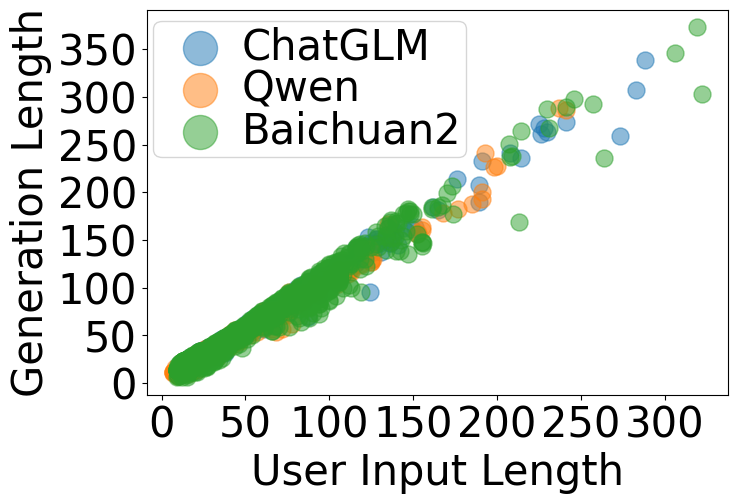}
		\caption{CT}
		\label{fig:corr_ct}
	\end{subfigure}
	\centering
	\begin{subfigure}{0.165\linewidth}
		\centering
		\includegraphics[width=1\linewidth]{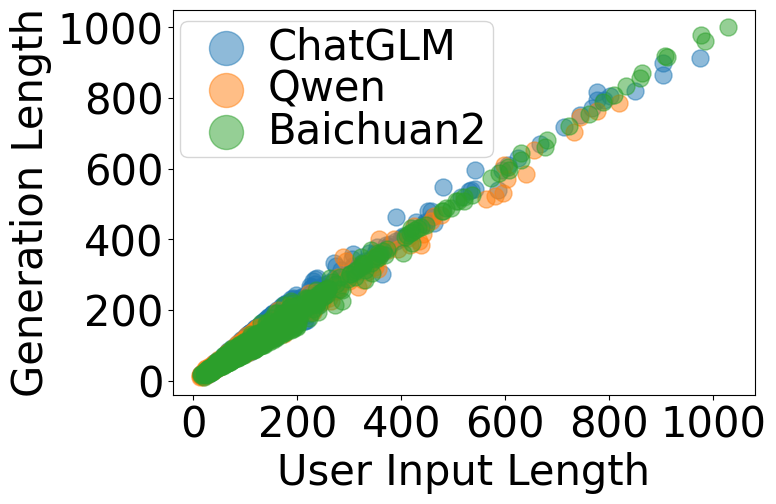}
		\caption{BF}
		\label{fig:corr_bf}
	\end{subfigure}
	\centering
	\begin{subfigure}{0.1675\linewidth}
		\centering
		\includegraphics[width=1\linewidth]{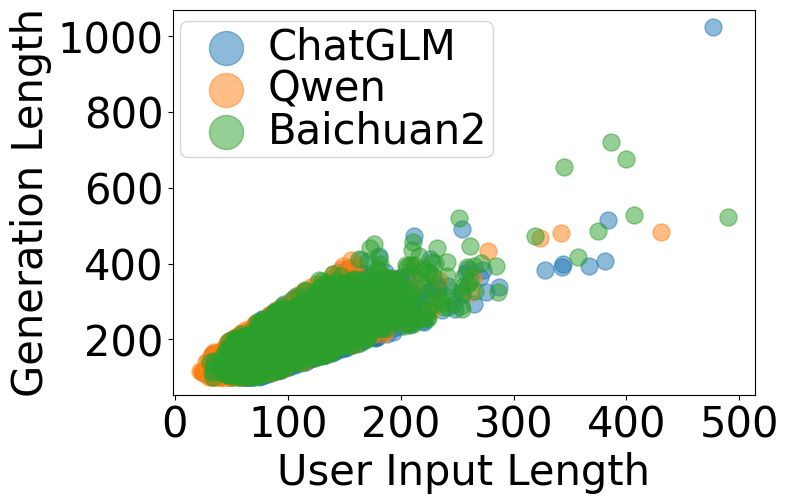}
		\caption{CC}
		\label{fig:corr_cc}
	\end{subfigure}
	\caption{Apps with a strong positive correlation between the user input length and the request generation length.}
	\label{fig:corr}
\end{figure*}

To solve the first problem, existing studies propose continuous batching, which can dynamically remove the completed requests and add newly arrived requests during the batch serving procedure, thus reducing the computational waste caused by request waiting. However, many LLM serving frameworks \cite{yu2022orca, deepspeed-fastgen} equipped with continuous batching leverage conservative GPU memory management strategies that limit the number of parallel-processing requests to avoid OOM errors and speedup inference, which fails to fully exploit the powerful parallel computing capability of GPUs and hurts throughput. 
To mitigate the second problem, previous studies try to compress LLMs using quantization \cite{yao2022zeroquant, xiao2023smoothquant,frantar2023optq, shao2023omniquant, chee2024quip} and pruning \cite{wang2020structured, xu2021rethinking, xu2022dense, liu2023deja, sun2023simple} to reserve more GPU memory to support a larger batch size, while other studies also compress the key-value cache \cite{ge2023model, zhang2024h2o, liu2024scissorhands}. However, they compromise the quality of generation and fail to reduce the computational waste.



\begin{table}[t]
	\centering
	\begin{tabular}{c|c|c|c|c|c|c}
		\hline
		& MT     & GC   & TD          & CT & BF &CC \\ \hline\hline
		ChatGLM-6B        & 0.967         & 0.981          & 0.778           & 0.996    & 0.992 & 0.771  \\ \hline
		Qwen-7B-Chat        & 0.962 & 0.991        & 0.853           & 0.995 & 0.992 & 0.768      \\ \hline
		Baichuan2-7B-Chat        & 0.996          & 0.987 & 0.829           & 0.990 & 0.995 & 0.809      \\ \hline
	\end{tabular}
	\caption{Pearson correlation coefficient between user input lengths and request generation lengths for various applications.}
	\label{tab:corr}
	\vspace{-0.5cm}
\end{table}

In this paper, we find that in the LMaaS scenario, there are many popular applications where the request generation length is positively correlated with the length of raw user input text, such as the multilingual machine translation (MT), grammar correction (GC), text detoxification (TD), code translation(CT), bug fixing (BF), and code comment(CC). 
To confirm this observation, we build 2,000 requests for each of the six applications from existing datasets \cite{wmt18, logacheva2022paradetox, stahlberg2021synthetic, lu2021codexglue, yasunaga2021break} and feed the requests to three LLMs ChatGLM-6B \cite{chatglm}, Qwen-7B-Chat\cite{qwen}, and Baichuan2-7B-Chat\cite{baichuan}. As shown in Fig. \ref{fig:corr}, for these applications, the length of user input texts and generated texts have a significant positive correlation. We also list the Pearson coefficients of user input length and request generation length in Table \ref{tab:corr}, where we can find that for various LLMs, the Pearson coefficients of most applications are greater than 0.8, indicating a strong positive correlation. Therefore, for these applications, the user input length can greatly help predict the generation length of requests. 

In addition, it is important to note that generation-length-predictable applications include not only those applications where the generation length is positively correlated with the user input length but also the applications where the generated texts for various requests have a similar or same generation length, such as LLM-based recommendation and classification, whose generation lengths can be accurately predicted with the average request generation length. In the LMaaS platform of Ant Group, more than 60\% of requests come from generation-length-predictable applications. Therefore, generation length prediction is applicable to most requests, and optimizing the serving efficiency for these requests can greatly improve the overall serving efficiency of the LMaaS platform.

In this paper, we mainly focus on the applications where
the user input length and the request generation length are
positively correlated and propose Magnus to predict the request generation length and leverage the predicted results to enable efficient batch serving in the LMaaS scenario from a pure scheduling perspective. Magnus is composed of four components, a generation length predictor, an adaptive batcher, a serving time estimator, and a batch scheduler. The generation length predictor leverages the user input length, application-level semantic features extracted from the instruction, and user-level semantic features extracted from the user input to predict the request generation length with a random forest regressor. 
The adaptive batcher uses the proposed wasted memory access (WMA) metric to model the computational waste in the batch serving procedure. With the goal of reducing WMA, the batcher can batch the requests with similar predicted generation lengths together and set an appropriate batch size for each batch, thus greatly mitigating the two problems of existing serving systems and achieving efficient batch serving.

To improve the QoS, when an LLM instance becomes idle, the serving time estimator uses the KNN regression to estimate the serving time of each queued batch using the batch size, request lengths, and predicted generation lengths of requests in the batch. Then, the batch scheduler selects a target batch for the LLM instance to serve based on the highest response ratio next (HRRN) policy, which reduces the request queuing time and hence decreases the request response time.

We implement a prototype system of Magnus and verify its superior performance under multi-application-synthetic workloads with 7 ChatGLM-6B instances deployed on 7 NVIDIA V100 32GB GPUs. Experiments confirm that compared to baselines such as continuous batching and model compression, Magnus can improve the serving efficiency with negligible overhead. It improves request throughput by 66\% to 234\% and reduces the average serving latency by 60.3\% to 89.7\% across various request arrival rates.

\section{Preliminaries and Motivation}
\label{sec:preliminaries}
\subsection{Autoregressive Generation for Language Models}
Language models receive a token sequence as input to output the probability distribution of the next token.  By sampling from the probability distribution, a new token can be obtained.  Thus, a sequence of tokens can be generated by iteratively concatenating the newly generated token after the input sequence and feeding it into the language model. This generation pattern is followed by LLMs and is called autoregressive generation. 

Autoregressive generation completes when the end-of-sequence (EOS) token is generated or the number of iterations reaches the preset maximal generation length limit. The process of generating one new token for a single request is called an iteration, and the number of generated tokens is called \textit{\textbf{request generation length}}, which is equal to the number of iterations. In the LMaaS scenario, requests are composed of the instruction and the user input. The length (i.e., number of tokens) of the user input text is called \textit{\textbf{user input length}} and the length of the whole request text is called \textit{\textbf{request length}}. 

By training on large-scale corpora and fine-tuning on instruction datasets, LLMs are capable of following instructions and completing specific tasks such as machine translation, grammar correction, and bug-fixing. For example, when ``Translate to German: I love you" is input as a request, the LLM can generate "Ich liebe dich EOS" in four iterations, thus completing the translation task.

Applications in the LMaaS scenario, such as machine translation and bug-fixing, focus more on the generation quality instead of diversity. Hence, greedy sampling and beam search \cite{freitag2017beam} are commonly adopted. Since beam search is computationally expensive, the LLM always generates tokens via greedy sampling where the token with the highest possibility is sampled as the next token in each iteration and hence the generated text is always identical for the same request.

\begin{figure}[t]
	\setlength{\belowcaptionskip}{-0.25cm} 
	\centering
	\includegraphics[width=0.9\linewidth]{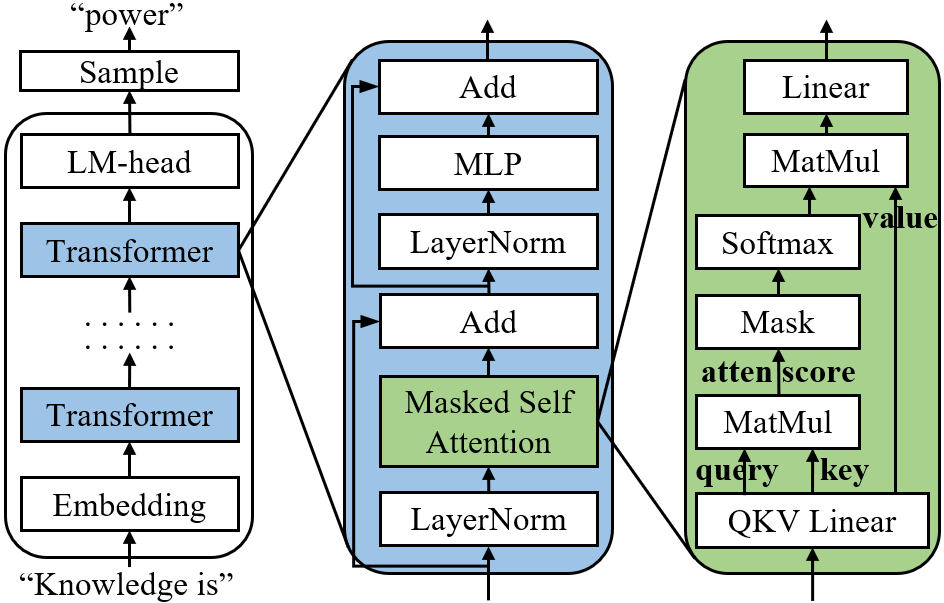}
	\caption{LLM inference procedure. The blue and green rounded rectangles depict the computation flow of the transformer block and masked-self attention, respectively.}
	\label{fig:llm_inference}
\end{figure}

\subsection{Inference Procedure for Large Language Models}
The LLM inference procedure is shown in Fig. \ref{fig:llm_inference}. After the text "Knowledge is" is fed to the LLM, the input token sequence is successively processed through a linear layer known as embedding, a stack of transformer blocks, a linear layer known as lm-head, and finally a possibility distribution is output from which ``power" is sampled as the next token. 

The most important part of the LLM is the transformer block, of which the masked self-attention module is the core component that elevates it \cite{vaswani2017attention}. In the inference procedure of the masked self-attention module shown in Fig. \ref{fig:llm_inference}, the input tensor of each token first passes through three different linear layers to derive the query, key, and value tensors, respectively. The query and key matrices are then multiplied to produce the attention score matrix, which represents the correlation between the tokens.
In order to make each token attend only to its preceding tokens, a masking operation is performed on the attention score matrix, where a very large negative number is added to the attention scores representing the correlation between tokens and their succeeding tokens. Therefore, these attention scores become negligible in the relevant matrix obtained after the soft-max operation. Next, by multiplying the relevant matrix with the value matrix, each token extracts features from its preceding tokens to update itself. Finally, these updated tensors are passed through a linear layer to obtain the output of the masked self-attention module.

\begin{figure}[t]
	\setlength{\belowcaptionskip}{-0.25cm} 
	\centering
	\includegraphics[width=0.888\linewidth]{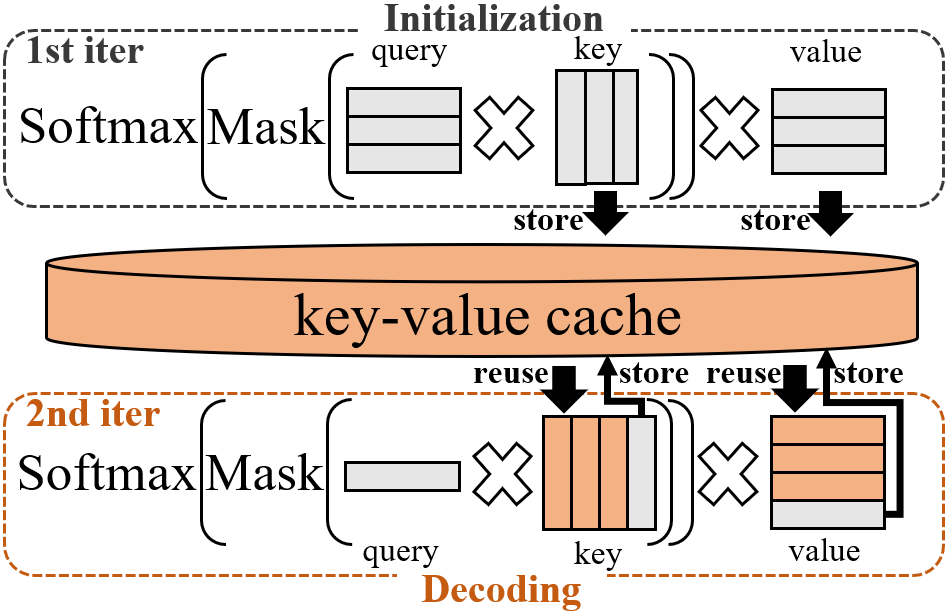}
	\caption{Key-value cache usage in two phases.  The request has 3 tokens and the gray and orange grids represent the newly derived and reused key and value tensors, respectively.}
	\label{fig:key_value_cache}
\end{figure}

\subsection{Key-Value Cache and Two-Phase Inference}
In a stateless implementation of autoregressive generation for LLMs, the request and all the generated tokens are concatenated to be fed into the LLM in each iteration, and hence the key and value tensors are recomputed across iterations, which results in a severe computational overhead. To avoid such recomputation, key, and value tensors are cached for reuse. Thus, the generation procedure can be divided into two phases, an initialization phase and a decoding phase. 

The first iteration is the \textit{\textbf{initialization phase}} where the request is fed to the LLM and the key and value tensors are computed and stored in the key-value cache. The subsequent iterations are in the \textit{\textbf{decoding phase}} where only the token generated in the last iteration is fed to the LLM, and the LLM computes the query, key, and value tensors for the input token, utilizes the key-value cache to complete the self-attention operation and store the newly computed key and value tensors in the cache. Fig. \ref{fig:key_value_cache} shows an example of the key-value cache usage in the first two iterations, where the request length of the request is 3.

\subsection{Batch Serving Procedure for Large Language Models}
\label{sec:llm_batch_inference_process}
\begin{figure}[t]
	\setlength{\belowcaptionskip}{-0.25cm} 
	\centering
	\includegraphics[width=0.9\linewidth]{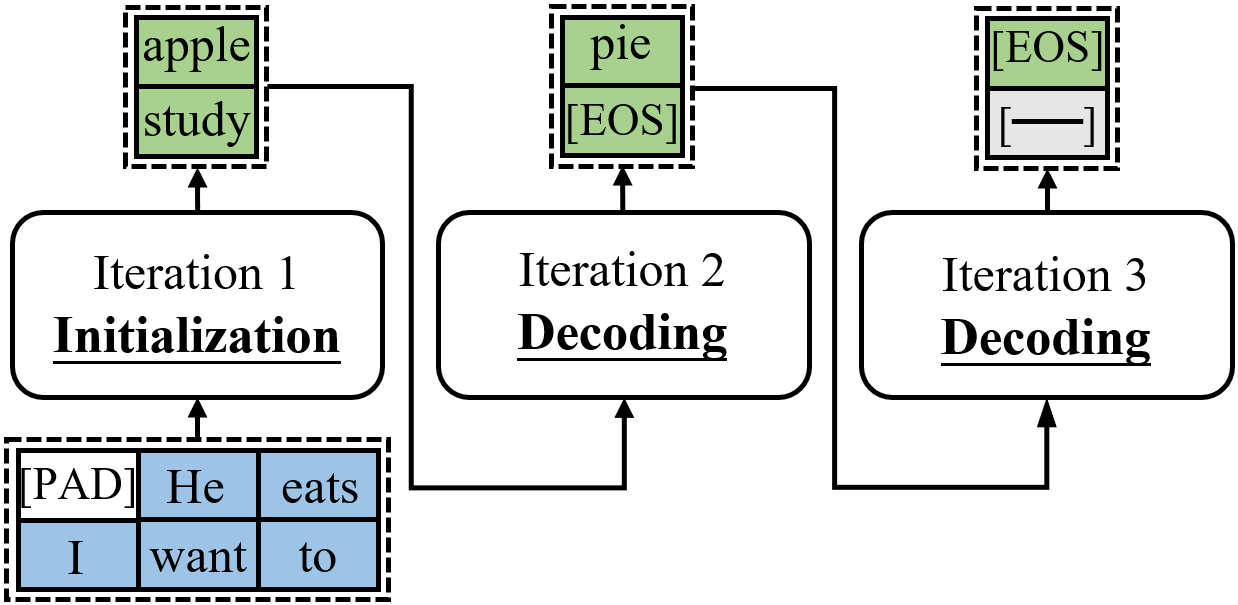}
	\caption{Batch serving procedure for LLMs. The blue and green grids represent tokens of the request and valid tokens generated by the LLM, respectively.}
	\label{fig:batch_inference}
\end{figure}

Batch serving is commonly exploited to improve the computing efficiency of neural network inference on GPUs. 
Fig. \ref{fig:batch_inference} illustrates the typical batch serving process of existing LLM inference engines such as huggingface-transformers \cite{wolf2020transformers} and deepspeed-inference \cite{aminabadi2022deepspeed}.

As shown in Fig. \ref{fig:batch_inference}, firstly, all the requests in a batch are padded to the same length as the longest one using the pad token. Next, the batch is fed to the LLM for autoregressive generation. To prevent the pad token from being attended by other tokens, the corresponding attention score is also masked. Until the request with the longest generation length in the batch finishes generation (i.e., the EOS token is sampled or the number of iterations reaches the preset maximal generation length limit), the generated results are returned together. Since the generation lengths of requests in a batch are diverse, early-completed requests continue generating until the batch serving completes, but the tokens generated after the EOS token will be ignored in the final response. This invalid generation process is called \textit{\textbf{request waiting}}. 

In this paper, we refer to the longest request length of requests in a batch as the \textit{\textbf{batch length}}, which is equal to the token sequence length of each request after padding. Moreover, we refer to the longest generation length of requests in a batch as the \textit{\textbf{batch generation length}}, which is equal to the total number of iterations in the batch serving procedure.

\subsection{Inefficiency of Vanilla Scheduling}
\label{sec:inefficiency_of_static_batching}
Production-grade inference serving systems such as TensorFlow Serving \cite{tensorflowserving} and Triton Inference Server \cite{triton},  leverage a fixed batch size to serve requests in an FCFS manner, causing a severe inefficiency for LLM inference. In this paper, we call such a scheduling method vanilla scheduling.

\textbf{Computational waste.} Since the requests are grouped into batches according to the arrival time, requests in the same batch not only have various request lengths but also have various request generation lengths. Different request lengths lead to padding. The pad tokens are involved in all computations during the batch serving procedure, but due to the mask operation, they do not contribute anything to the generated results, thus wasting computation. In addition, different request generation lengths lead to request waiting, where the requests that complete generation early not only cannot be returned immediately but also continue generating invalid tokens, which further exacerbates computational waste.

\textbf{Small batch size.} Not only tokens of the raw request text and generated valid tokens but also the pad tokens and generated invalid tokens produce key and value tensors that are cached in the GPU memory. Therefore, for a batch, the longer the batch length and the batch generation length are, the larger the key-value cache is, and the more GPU memory is consumed. As the vanilla scheduling lacks perception of the request length and request generation length, it often assumes that all requests have the preset maximal request length and maximal request generation length. Thus, to avoid OOM errors, vanilla scheduling sets the batch size to 
\begin{equation}
	\label{eq:static_batch_size}
	\beta=\lfloor \frac{\Theta}{(L_{max} + G_{max}) \cdot \Delta} \rfloor,
\end{equation}
where $\Theta$ denotes the available GPU memory to store the key-value cache, $\Delta$ represents the memory usage of key and value tensors of one token, which is determined by the LLM's architecture, $L_{max}$ and $G_{max}$ symbolize the preset maximal request length and maximal request generation length, respectively. Since $L_{max}$ and $G_{max}$ are large, $\beta$ is small and prevents GPUs from being fully utilized.

\begin{figure}[t]
	\centering
	\begin{subfigure}{1\linewidth}
		\centering
		\includegraphics[width=0.9\linewidth]{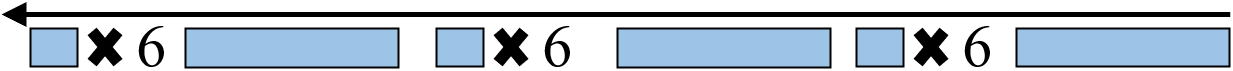}
		\caption{Request arrival.}
		\label{fig:prompt_arrival}
	\end{subfigure}
	\setlength{\belowcaptionskip}{-0.2cm}
	\centering
	\begin{subfigure}{0.49\linewidth}
		\centering
		\includegraphics[width=0.875\linewidth]{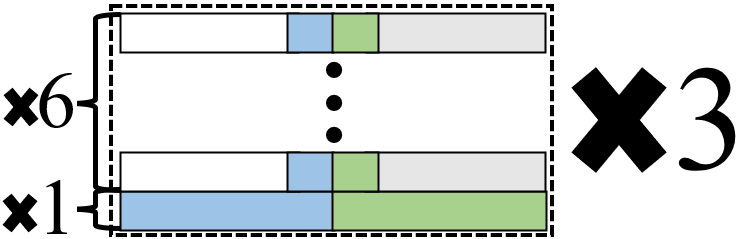}
		\caption{Vanilla scheduling.}
		\label{fig:static_batching}
	\end{subfigure}
	\centering
	\begin{subfigure}{0.49\linewidth}
		\centering
		\includegraphics[width=0.975\linewidth]{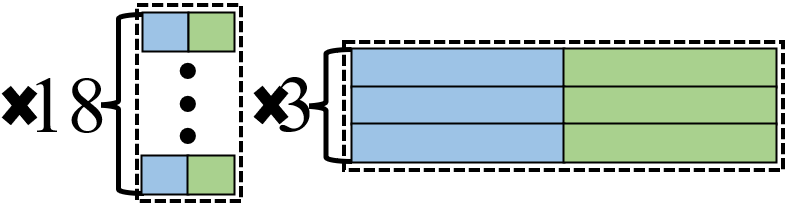}
		\caption{Magnus.}
		\label{fig:predbatch_batching}
	\end{subfigure}
	\caption{Magnus vs. Vanilla scheduling. After 21 requests arrive, Magnus separately batches the requests with small length and generation length, and the requests with large length and generation length into 2 batches, while vanilla scheduling batches them into 3 batches with a fixed batch size of 7 according to the arrival order. Hence, Magnus reduces the total serving time by 75.2\% compared with vanilla scheduling. The white and gray grids represent the pad tokens and invalid generated tokens, respectively.}
	\label{fig:PredBatch_vs_static batching}
\end{figure}

\textbf{Case study.} To confirm the inefficiency of vanilla scheduling, we conduct an experiment with ChatGLM-6B deployed on an NVIDIA V100 32GB GPU, where the "large" requests whose request length and generation length are about 1000, and "small" requests whose request length and generation length are about 10 arrive in the order shown in Fig. \ref{fig:prompt_arrival}. In the experiment, we leverage huggingface-transformers as the inference engine to load and run the LLM. Vanilla scheduling batches them in an FCFS manner with a fixed batch size of 7, which leads to a total serving time of 242s. However, as shown in Fig. \ref{fig:predbatch_batching}, Magnus groups the small requests and large requests into two batches with batch sizes of 18 and 3, respectively, and the total serving time is 60s, greatly outperforming vanilla scheduling.

\section{Solution Description}
\label{sec:system_design}
Magnus is proposed to mitigate the computational waste and enlarge the batch size via generation length prediction, thus enabling efficient batch serving for LMaaS.

\subsection{Magnus Overview}
Magnus is composed of four components, a generation length predictor, a WMA-directed adaptive batcher, a serving time estimator, and an HRRN batch scheduler.

\begin{figure}[t]
	\setlength{\belowcaptionskip}{-0.5cm} 
	\centering
	\includegraphics[width=1\linewidth]{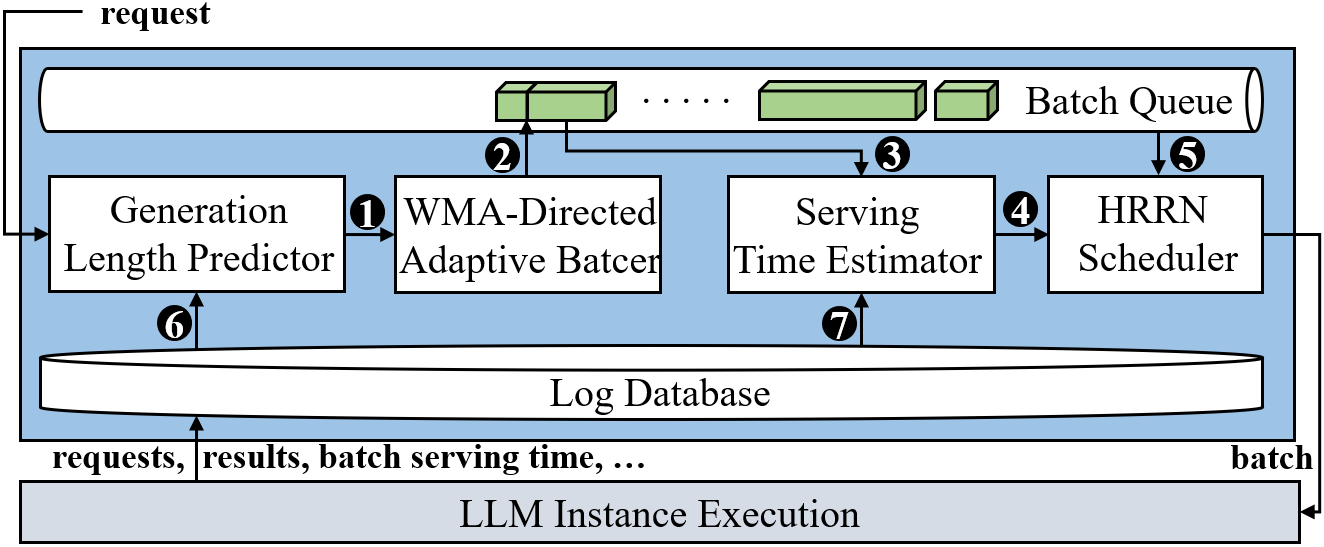}
	\caption{Magnus workflow. The arrows represent data transferred between modules.}
	\label{fig:system_overiew}
\end{figure}

As shown in Fig. \ref{fig:system_overiew}, when a request arrives, the generation length predictor predicts its generation length and sends \ding{182} the request and the predicted generation length to the adaptive batcher. The batcher inserts \ding{183} the request to a queued batch whose requests have a similar length and predicted generation length, thus mitigating computational waste. When an LLM instance finishes generation, the serving time estimator will estimate the batch serving time for queued batches according to \ding{184} the batch length, predicted batch generation length, and batch size. Then, the HRRN batch scheduler uses the \ding{185} estimated batch serving time to schedule \ding{186} a queued batch to the LLM instance using the HRRN policy which prioritizes the batches by trading off their queuing time and inference time. In addition, Magnus periodically leverages newly collected \ding{187} request information such as requests and their generation lengths, and \ding{188} batch information such as batch size, batch length, batch generation length, and batch serving time from the log database to refine the generation length predictor and the serving time estimator with continuous learning. 

\subsection{Generation Length Predictor}
The generation length predictor takes the user input length, the application-level semantics extracted from the instruction, and the user-level semantics extracted from the user input to predict the request generation length. 

In the LMaaS scenario, the request is composed of the instruction and the user input. As we have verified in Section \ref{sec:introduction},  many popular applications have a positive correlation between the user input length and request generation length. Hence, the user input length is of great significance for the predictor to learn the request generation length. 

However, not only different applications, but even different tasks of the same application may have different correlations between the user input length and request generation length, which poses a challenge for effective predicting.
For example, for the bug-fixing application, the request generation length and user input length are similar, whereas, for the code comment application, the request generation length is always longer than the user input length. Besides, for the code translation application, when translating C++ code to Python code, the request generation length is always shorter than the user input length due to the concise syntax rules of Python, while for the task of translating Python code to C++ code, the request generation length is often longer than the user input length. 
Since the number of tasks can be very large, it is expensive to train and run a customized predictor for each task. To make the predictor learn the correlation between the user input length and request generation length for various applications and tasks, we exploit semantic features of the instruction to help the predictor distinguish different applications and their tasks because the applications and their tasks are identified by the instruction.

Moreover, motivated by the GPTCache \cite{bang2023gptcache}, with the same instruction, semantically similar user inputs tend to have similar generated responses and hence have similar request generation lengths. Therefore,  the predictor can make use of the user input to further improve the prediction accuracy based on the similarity between the user input semantics.

\begin{figure}[t]
	\setlength{\belowcaptionskip}{-0.5cm} 
	\centering	\includegraphics[width=0.85\linewidth]{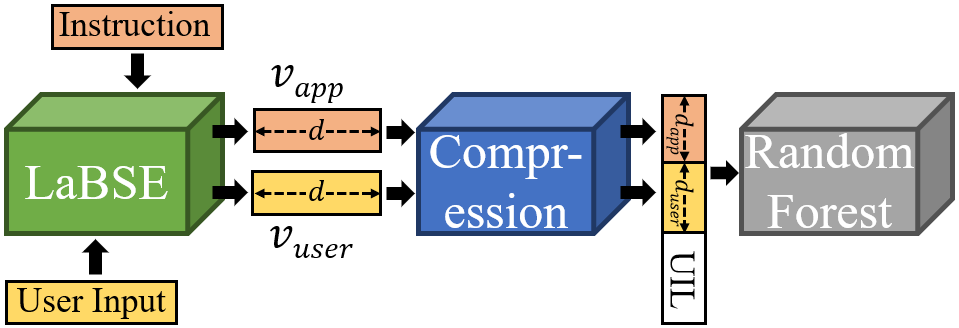}
	\caption{The architecture of the generation length predictor, where UIL is short for the user input length.}
	\label{fig:generation_length_predictor}
\end{figure}

Based on the above analysis, we design the generation length predictor for Magnus, which leverages the user input length, instruction text, and user input text as input to predict the request generation length. 
The generation length predictor is composed of a language-agnostic BERT sentence embedding (LaBSE) \cite{feng2022language}, a compression module, and a random forest regressor.  We provide a simplified high-level overview of the generation length predictor in Fig. \ref{fig:generation_length_predictor}. Firstly, the LaBSE module takes the user input and instruction as input to extract the user-level and application-level semantic features, respectively, and produces two embedding vectors $v_{user}, v_{app}\in \mathbb{R}^{d}$, where $d=768$. Then, to control the complexity of the regressor, $v_{user}$ and $v_{app}$ are compressed with the compression module which evenly divides them into $d_{user}$ and $d_{app}$ groups with group size $\frac{d}{d_{app}}$ and $\frac{d}{d_{user}}$, and aggregates each group into a single value by summing the in-group values. Moreover, the sum is divided by the square root of the group size to ensure numerical stability. Finally, the compressed embedding vectors are concatenated with the user input length to feed to the random forest regressor to predict the generation length. Through several attempts, we set $d_{user}$ and $d_{app}$ to 16 and 4, separately, to achieve a high prediction accuracy.

We validate the effectiveness of the generation length predictor design for three LLMs on eight tasks of six applications (see Section \ref{sec:experiments}). For each task, we construct 2,000 requests as the training set and 500 requests as the test set. Then, we evaluate the prediction error of the following four strategies. (1) User Input Length Only (\textit{\textbf{UILO}}), which directly utilizes the user input length as predicted request generation length. (2) Random Forest (\textit{\textbf{RAFT}}), which trains eight random forest regressors for the eight tasks to predict the request generation length merely using the user input length. (3) Instruction (\textit{\textbf{INST}}), which trains a random forest regressor to predict the request generation length for all the tasks using the compressed application-level semantic features and user input length. (4) User Input (\textit{\textbf{USIN}}), which adds the compressed user-level semantic features to inputs on the basis of INST. From Table \ref{tab:predicting_methods}, we can find that the prediction error of directly using the user input length as the predicted result is high, and the random forest regressor can capture the correlation between the user input length and the request generation length. The application-level semantics can help the regressor to distinguish between different application tasks while the user-level semantics can further improve the prediction accuracy.


\begin{table}[t]
	\centering
	\begin{tabular}{c|c|c|c|c}
		\hline
		& UILO     & RAFT   & INST          & USIN \\ \hline\hline
		ChatGLM-6B        & 33.961         & 16.158          & 16.156           & 15.649      \\ \hline
		Qwen-7B-Chat        & 37.106 & 14.005        & 14.267           & 13.318      \\ \hline
		Baichuan2-7B-Chat        & 37.814          & 16.896 & 16.871           & 16.018      \\ \hline
	\end{tabular}
	\caption{Root Mean square error (RMSE) of various predicting methods. RMSE represents the average gap between the predicted and actual request generation length.}
	\label{tab:predicting_methods}
	\vspace{-0.5cm}
\end{table}

During prediction, the LaBSE module packs the instruction and user input into a batch for processing to improve computing efficiency. When the workload is heavy, multiple generation length predictors can be deployed to make predictions parallelly, thus improving the throughput of prediction.

To accommodate the user input variations, the generation length predictor is steadily enhanced through continuous learning. Every 3 minutes, Magnus collects log data of the newly served requests whose prediction error is greater than 10 tokens and exceeds 10\% of the actual request generation length, augments the train set with these collected data, and retrains the random forest regressor. Continuous learning is asynchronous to the online prediction and does not affect the prediction speed. 

\subsection{WMA-Directed Adaptive Batcher}
The adaptive batcher inserts requests to the queued batches based on their lengths and the predicted request generation lengths with the goal of reducing computational waste.

Since the major overhead of LLM batch serving comes from GPU memory access \cite{dao2022flashattention}, we propose the wasted memory access (WMA) metric to model computational waste during batch serving, which is equal to the number of times that a token's key and value tensors are read but do not contribute anything to the generated result.

\begin{figure}[t]
	\setlength{\belowcaptionskip}{-0.5cm} 
	\centering	\includegraphics[width=0.9\linewidth]{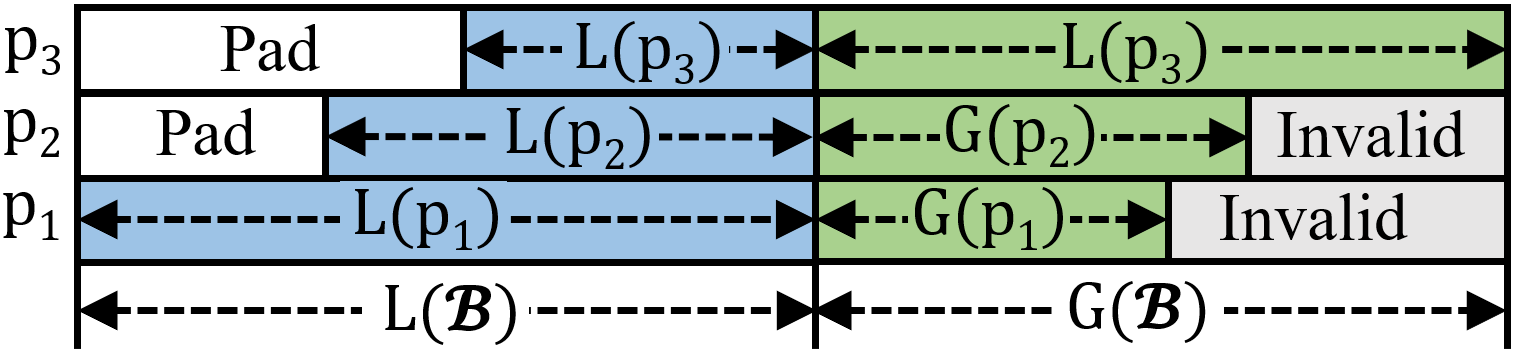}
	\caption{Illustration of the notations utilized for calculating WMA. The blue, green, white, and gray grids represent tokens of raw requests, valid tokens generated by the LLM, pad tokens, and generated invalid tokens, respectively.}
	\label{fig:wma}
\end{figure}

In the batch serving procedure, the WMA of a request is composed of two parts. The first part $WMA_{gen}$ is caused by the pad tokens before the EOS token is generated, and the second part $WMA_{wait}$ comes from the invalid tokens generated during the request waiting.  Fig. \ref{fig:wma} illustrates the notations utilized to calculate the WMA for a batch composed of three requests with various request lengths and request generation lengths.

The length of a batch $\mathcal{B}$ is defined as $L(\mathcal{B}) = \max_{p\in \mathcal{B}} L(p)$ and the batch generation length is defined as $G(\mathcal{B}) = \max_{p\in \mathcal{B}} G(p)$, where $L(p)$ and $G(p)$ represent the length and generation length of the request $p$. For a request $p\in \mathcal{B}$, its number of pad tokens is $L(\mathcal{B})-L(p)$. The key and value tensors of these pad tokens are read for computing in each iteration, but they do not have any contribution to the generated text. Therefore, these memory accesses are wasted. Until the EOS token is generated, key and value tensors of these pad tokens are read $G(p)$ times, and thus the $WMA_{gen}$ of $p$ is calculated by
\begin{equation}
	\label{eq: prompt_gen_wma}
	WMA_{gen}(p) = G(p) * (L(\mathcal{B})-L(p)).
\end{equation}

Besides, after the EOS token is generated, $p$ enters the waiting phase, where the padded request and all previously generated tokens are read, and computed, and newly produced key and value tensors are cached in each iteration. Since tokens generated in the waiting phase are invalid and will be ignored in the final response, these memory accesses are wasted as well, which causes a $WMA_{wait}$ of 
\begin{equation}
	\label{eq:prompt_wait_wma}
	WMA_{wait}(p)=\sum\limits_{g=G(p)}^{G(\mathcal{B})}(g + L(\mathcal{B})).
\end{equation}

The WMA of a batch $\mathcal{B}$ is defined as the maximal WMA of all its requests, which is expressed by 
\begin{equation}
	\label{eq:batch_wma}
	WMA(\mathcal{B})=\max\limits_{p\in\mathcal{B}} (WMA_{gen}(p) + WMA_{wait}(p)).
\end{equation}

When the adaptive batcher receives a request $p$, it iterates batches in the waiting queue,  substitutes $G(p)$ with the predicted generation length  $G'(p)$ to compute the WMA of each batch after $p$ is inserted into it, and records the minimum WMA $\phi$ as well as the corresponding batch $\mathcal{B}_{\phi}$. To reduce the computational waste during batch serving, if $\phi$ is less than the given threshold $\Phi$, the request is inserted to $\mathcal{B}_{\phi}$, otherwise, a new batch is created with the request and inserted in the waiting queue.

Since the key-value cache is produced and consumes a lot of GPU memory during the batch serving, to avoid the OOM error, it is significant to prevent the batch size from being too large. For a batch $\mathcal{B}$, the memory consumed by the key-value cache is calculated by 
\begin{equation}
	\label{eq:batch_key_value_cahce_memory}
	MEM(\mathcal{B})=\beta\cdot (L(\mathcal{B}) + G(\mathcal{B})) \cdot \Delta,
\end{equation}
where $\beta$ is the batch size of $\mathcal{B}$, and $\Delta$ represents the memory usage of key and value tensors of one token. Based on the predicted generation length of requests in $\mathcal{B}$, the memory consumption of $\mathcal{B}$ can be estimated before serving. 

If the adaptive batcher finds that the estimated memory usage of batch $\mathcal{B}$ exceeds the available memory size $\Theta$ after inserting $q$ into $\mathcal{B}$, it skips $\mathcal{B}$ to prevent $p$ from being inserted into $\mathcal{B}$ and hence reduces OOM errors.

Therefore, the batches with small lengths and generation lengths can have a large batch size to take full advantage of the powerful parallel computing capability of GPUs, and the batches with large lengths and generation lengths can have a proper batch size to utilize GPUs as much as possible while avoiding the OOM error. Pseudo-code of the WMA-directed adaptive batching procedure is provided in Algorithm \ref{alg:batching}.

However, OOM errors still occur because there is a deviation between the estimated and actual memory usage due to the generation length prediction error. To solve the problem, Magnus evenly splits the batch that causes the OOM error into two small batches, sets these two batches uninsertable, and puts them back into the waiting queue. Since the batch size is halved, the probability of these two batches causing the OOM error is greatly reduced.

\subsection{Serving Time Estimator}
The serving time estimator estimates the LLM inference time for the batches waiting in the queue when an LLM instance becomes idle so that the HRRN batch scheduler can leverage the estimated results to determine the order of batch execution, thus reducing the request response time. 

According to the LLM batch serving procedure described in Section  \ref{sec:llm_batch_inference_process}, batches with similar length, generation length, and batch size have a similar number of iterations and a similar amount of memory accesses and computation in each iteration, thus having similar batch serving time. Therefore, KNN regression is naturally leveraged to predict the batch serving time of a batch based on its batch size, length, and generation length. 
When an LLM instance finishes generation and becomes idle, the estimator estimates the serving time for all the queued batches via KNN regression, where the maximal predicted generation length of all the batched requests is utilized as the generation length of the batch. 

Similar to the generation length predictor, the serving time estimator is also enhanced over time via continuous learning.  Every 2 minutes, Magnus collects log data of newly served batches and re-predicts their serving time with the actual generation length. Then, the length, generation length, and batch size data of batches whose prediction error is greater than 2 seconds and exceeds 20\% of the actual batch serving time are added to the train set to retrain the KNN regression model. The retraining is asynchronous to the estimating, which does not affect the online estimation.

\begin{algorithm}[t]
	\caption{WMA-Directed Adaptive Batching}\label{alg:batching}
	\KwIn{$\mathcal{Q}$: batch queue; $p$: request to be inserted; 
		\newline $\Theta$: available memory size; $\Phi$: WMA threshold;}
	$\phi\longleftarrow  +\infty$
	
	$\mathcal{B}_{\phi}\longleftarrow  \emptyset$
	
	\tcc{Find the batch with minimal WMA}
	
	\For{$\mathcal{B}$ in $\mathcal{Q}$}
	{
		$\mathcal{B}'\longleftarrow \mathcal{B}\cup \{p\}$
		
		\tcc{Memory consumption and WMA are computed with the predicted request generation length}
		
		\If{$MEM(\mathcal{B}')\le\Theta$ and $WMA(\mathcal{B}')<\phi$}
		{
			$\phi\longleftarrow WMA(B)'$
			
			$\mathcal{B}_{\phi}\longleftarrow\mathcal{B}$
		}
		
	}
	
	\If{$\phi<\Phi$}
	{
		$\mathcal{B}_{\phi}\longleftarrow\mathcal{B}_{\phi}\cup\{p\}$ \tcp{Insert $p$ into $\mathcal{B}_{\phi}$}
	}
	\Else
	{
		Create a new batch with $p$ and insert it in $\mathcal{Q}$
	}	
\end{algorithm}

\subsection{HRRN Batch Scheduler}
When an LLM instance finishes batch serving and becomes idle, the HRRN batch scheduler selects a target batch from the waiting queue according to its queuing time and estimated batch serving time, and then schedules it to the LLM instance.

The response ratio of a batch $\mathcal{B}$ is calculated by $\frac{T_q(\mathcal{B})}{T_s(\mathcal{B})}$, where $T_s(\mathcal{B})$ is the serving time of $\mathcal{B}$, which is replaced by the estimated batch serving time when calculating the response ratio, and $T_q(\mathcal{B})$ is the queuing time of $\mathcal{B}$, which is defined as the longest queuing time of requests in $\mathcal{B}$. The HRRN batch scheduler schedules the batch with the highest response ratio to be executed, which not only prioritizes batches with a short serving time to reduce the average queuing time of all batches but also prevents batches with a long serving time from waiting in the queue for too long, thus reducing the average request response time and improving the QoS. 

\subsection{Implementation}
We implement Magnus using Python 3.8 and libraries including PyTorch \cite{goues2019automated} as the base ML framework, sklearn \cite{pedregosa2011scikit} for implementing regressors, Sentence Transformers \cite{reimers2019sentence} and HuggingFace Transformers \cite{wolf2020transformers} for the NLP toolbox and implementations of the LaBSE and LLMs. 

On the control side, the four components of Magnus can be deployed in a distributed manner, and they communicate with each other through REST APIs. In addition, all four components are implemented as stateless microservices so that resources can be dynamically allocated to each component based on its performance and resource utilization.
On the inference side, each LLM instance runs in a separate worker process. When an OOM error occurs, the worker process will send the corresponding message to Magnus, empty GPU memory, and reload the LLM. The execution logs of all worker processes are stored in the Redis database and persist periodically. The separation of control and compute ensures the scalability of Magnus, as the components of Magnus and LLM instances can communicate through the network and do not have to run on the same physical machine.

\section{Evaluation}
\label{sec:experiments}

\begin{figure*}[t]
	\setlength{\belowcaptionskip}{-0.25cm} 
	\centering
	\begin{subfigure}{0.3375\linewidth}
		\centering
		\includegraphics[width=1\linewidth]{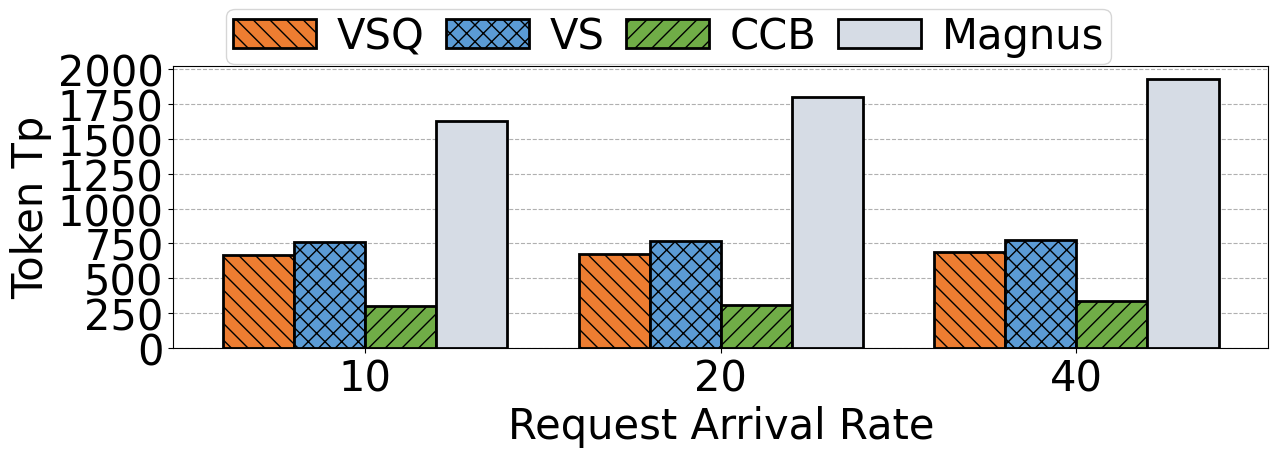}
		\caption{Token throughput.}
		\label{fig:eval_token_tp}
	\end{subfigure}
	\centering
	\begin{subfigure}{0.33\linewidth}
		\centering
		\includegraphics[width=1\linewidth]{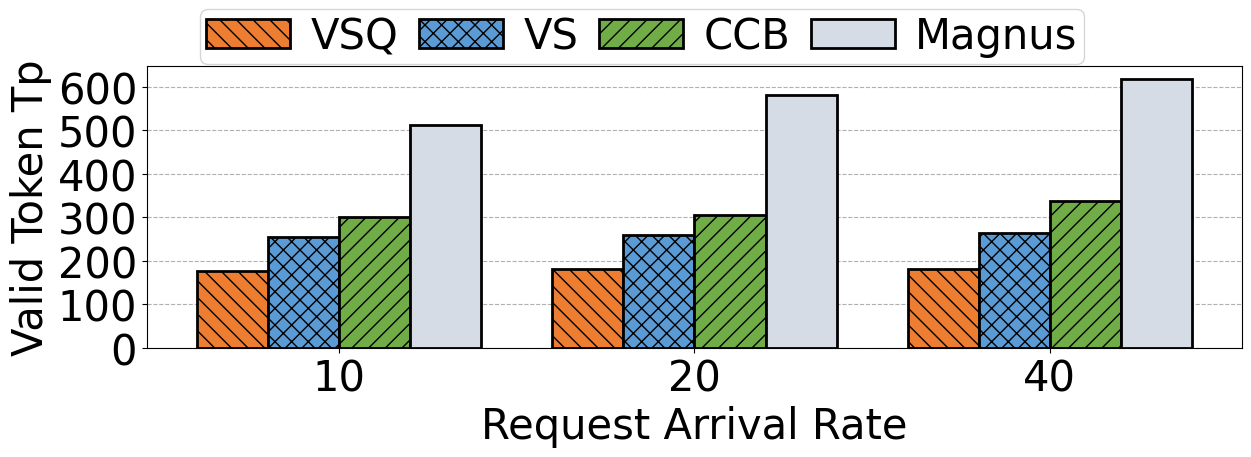}
		\caption{Valid token throughput.}
		\label{fig:eval_valid_token_tp}
	\end{subfigure}
	\caption{Token-level performance: token throughput (Tp) and valid token Tp under various request arrival rates.}
	\label{fig:eval_token}
\end{figure*}

\begin{figure*}[t]
	\setlength{\belowcaptionskip}{-0.25cm} 
	\centering
	\begin{subfigure}{0.32\linewidth}
		\centering
		\includegraphics[width=1\linewidth]{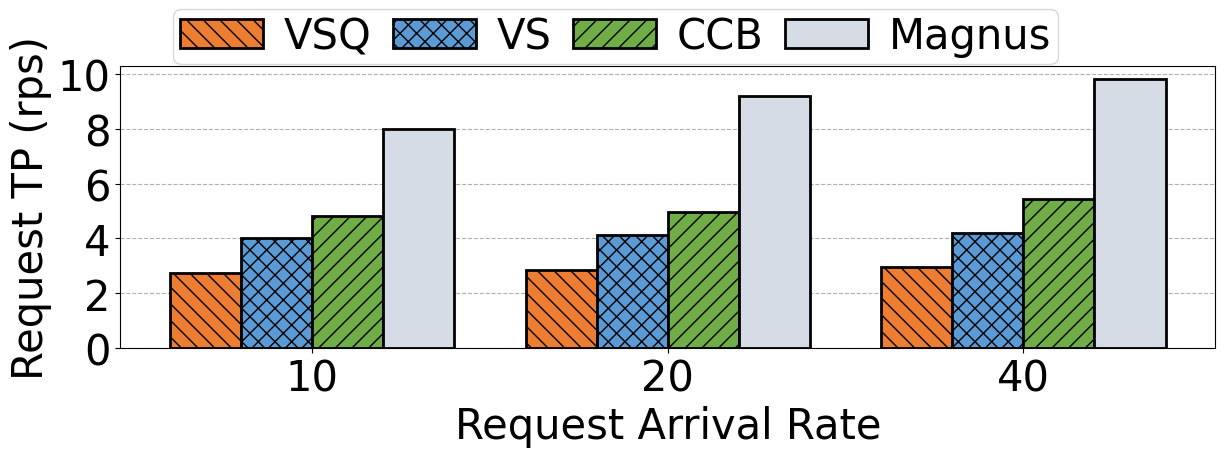}
		\caption{Request throughput.}
		\label{fig:eval_request_tp}
	\end{subfigure}
	\centering
	\begin{subfigure}{0.33\linewidth}
		\centering
		\includegraphics[width=1\linewidth]{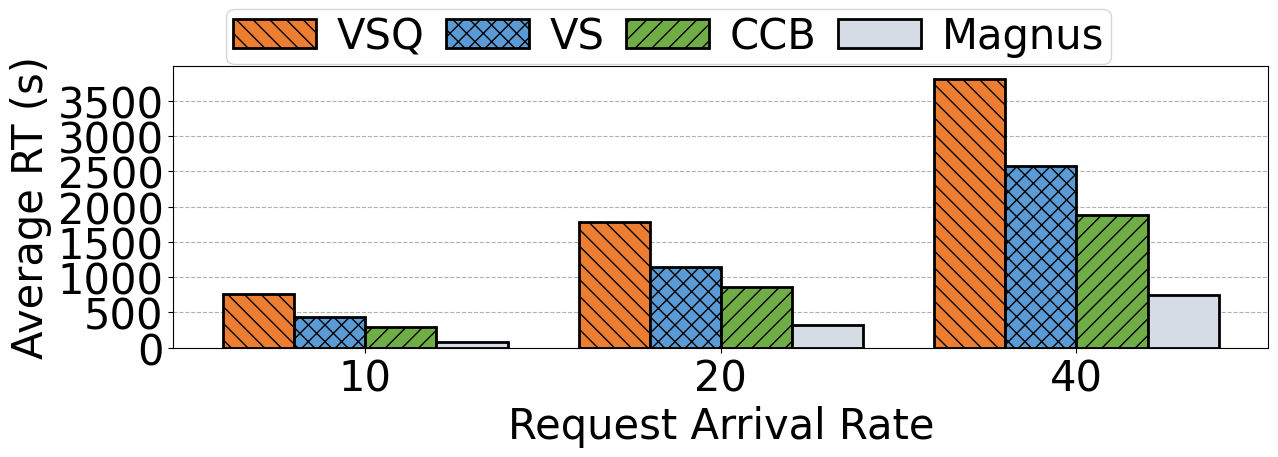}
		\caption{Average request response time.}
		\label{fig:eval_tail_res_time}
	\end{subfigure}
	\centering
	\begin{subfigure}{0.328\linewidth}
		\centering
		\includegraphics[width=1\linewidth]{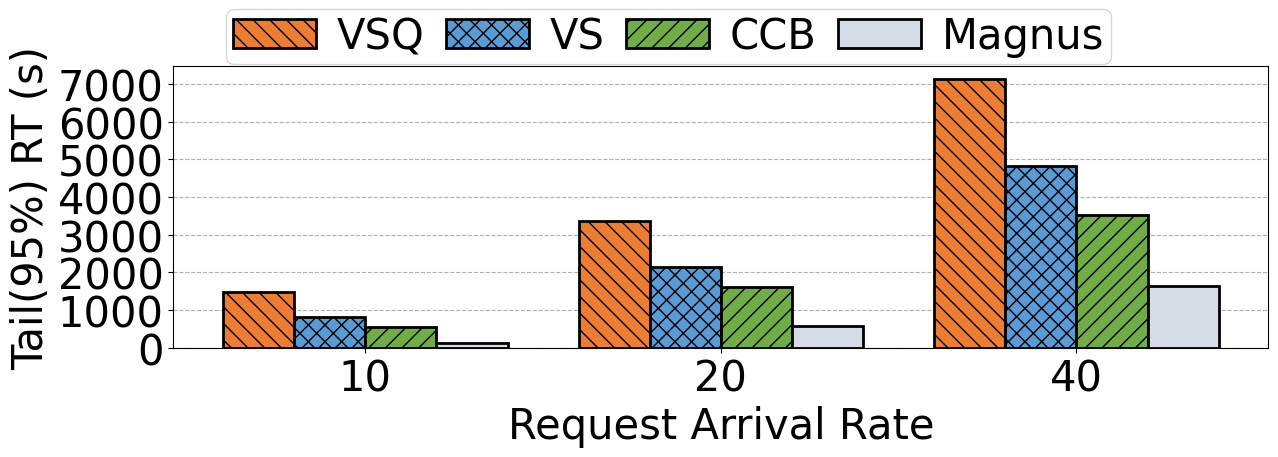}
		\caption{Tail (95\%) request response time.}
		\label{fig:eval_ave_res_time}
	\end{subfigure}
	\caption{Request-level performance: request throughput (Tp) and response time (RT) under various request arrival rates.}
	\label{fig:eval_request}
\end{figure*}

\subsection{Experiment Setup}
\noindent\textbf{Testbed}. Experiments are conducted on our testbed equipped with 8 NVIDIA V100 32GB GPUs connected over NVLink, 256GB CPU memory, and a 2.50GHz 24-Core Intel Xeon Platinum 8163 CPU.

\noindent\textbf{Applications, tasks, and datasets}. We synthesize a multi-application dataset of six applications including MT \cite{stahlberg2020neural}, GC \cite{wang2021comprehensive}, TD \cite{dale2021text}, CT \cite{zhu2022multilingual}, BF \cite{goues2019automated}, and CC \cite{codecomment} with existing datasets \cite{wmt18, logacheva2022paradetox, stahlberg2021synthetic, lu2021codexglue, yasunaga2021break}. Both MT and CT applications have two tasks, and thus there are a total of eight tasks. For each task, we randomly select 10,000 pieces of data from the data set to construct requests, where 7,500 requests are utilized to generate workloads, and the remaining 2,500 requests are leveraged to train the generation length predictor and serving time estimator of Magnus.

\noindent\textbf{Language model}. We evaluate a series of open-source LLMs, including ChatGLM-6B\cite{chatglm}, Qwen-7B-Chat \cite{qwen}, Baichuan2-7B-Chat\cite{baichuan}, LLaMA2-7B-Chat, and LLaMA2-13B-Chat \cite{llama2} via subjective testing, and select ChatGLM-6B as the LLM used in experiments because it can follow instructions well and has a relatively small model size. Hence, it can support a larger batch size to achieve a higher computing efficiency. 

\noindent\textbf{Settings}. The LLM generates tokens via greedy sampling. The maximum request length limit and the maximum request generation length limit are both set to 1024. 

\noindent\textbf{Workloads}. The arrival time of each request is determined by a Poisson distribution parameterized by the request rate.

\noindent\textbf{Metrics}. We use the request throughput, request response time, token throughput, and valid token throughput as evaluation metrics. The request response time is defined as the time between the request arrival and the return. The token throughput is defined as the number of tokens generated per second, including the invalid tokens generated after the EOS token.

\noindent\textbf{Baselines}. We compare Magnus with three baselines. Due to memory fragmentation, a considerable amount of memory can not be allocated \cite{kwon2023efficient}. Therefore, $\Theta$, the GPU memory size available to store the key-value cache is set to 70\% of the total GPU memory size minus the LLM size to mitigate OOM errors for Magnus and the baselines. 
\begin{itemize}
	\item \textbf{Vanilla Scheduling (VS)}: It serves requests in an FCFS manner with a fixed batch size of 7, which is calculated by Eq. (\ref{eq:static_batch_size}).
	\item \textbf{Vanilla Scheduling with 4-bit Quantization (VSQ)}: It serves requests in the same way as VS and compresses the LLM with 4-bit quantization, enabling a larger batch size of 10.
	\item \textbf{Conservative Continuous Batching (CCB)}: It dynamically removes finished requests and adds newly arrived requests for serving. We implement CCB for ChatGLM-6B with pytorch, where requests being served need to wait for the newly joined request to complete the initialization phase. For CCB, the number of parallel-processing requests is limited to 7 to avoid OOM errors.
\end{itemize}

\begin{figure*}[t]
	\setlength{\belowcaptionskip}{-0.25cm} 
	\centering
	\begin{subfigure}{0.3375\linewidth}
		\centering
		\includegraphics[width=1\linewidth]{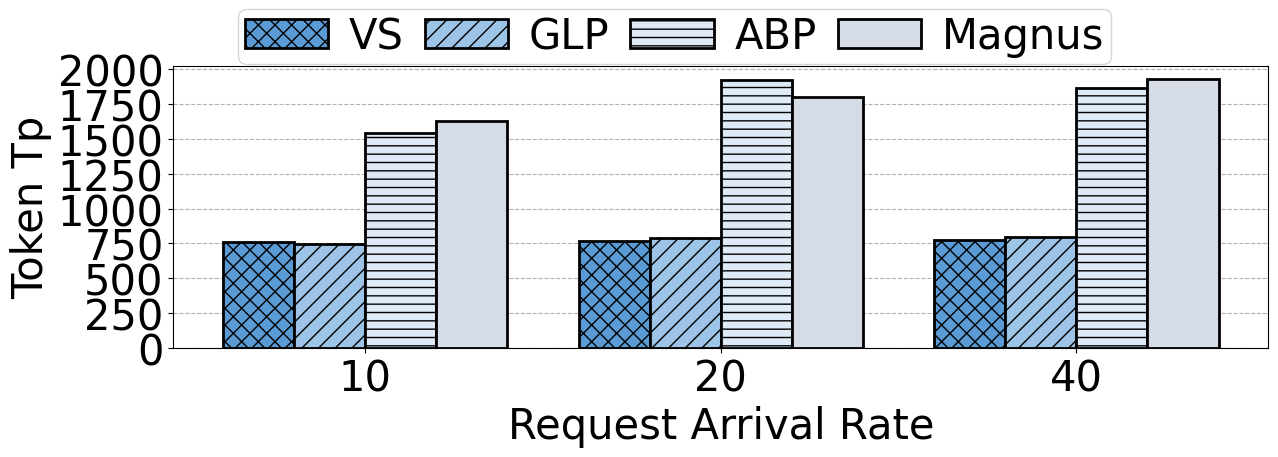}
		\caption{Token throughput.}
		\label{fig:abla_token_tp}
	\end{subfigure}
	\centering
	\begin{subfigure}{0.33\linewidth}
		\centering
		\includegraphics[width=1\linewidth]{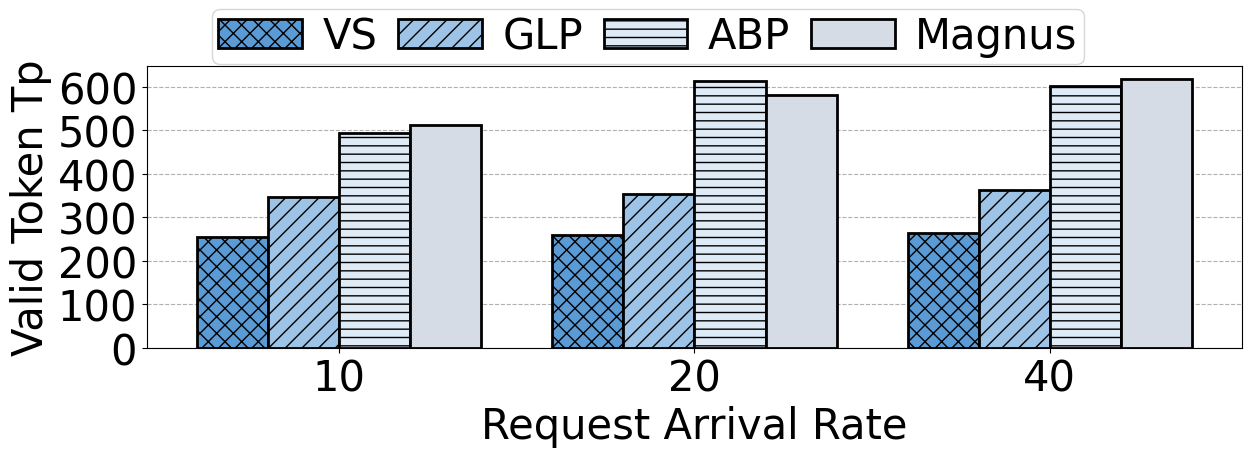}
		\caption{Valid token throughput.}
		\label{fig:abla_valid_token_tp}
	\end{subfigure}
	\caption{Token-level performance benefits gained from each component of Magnus.}
	\label{fig:abla_token}
\end{figure*}

\begin{figure*}[t]
	\setlength{\belowcaptionskip}{-0.25cm} 
	\centering
	\begin{subfigure}{0.32\linewidth}
		\centering
		\includegraphics[width=1\linewidth]{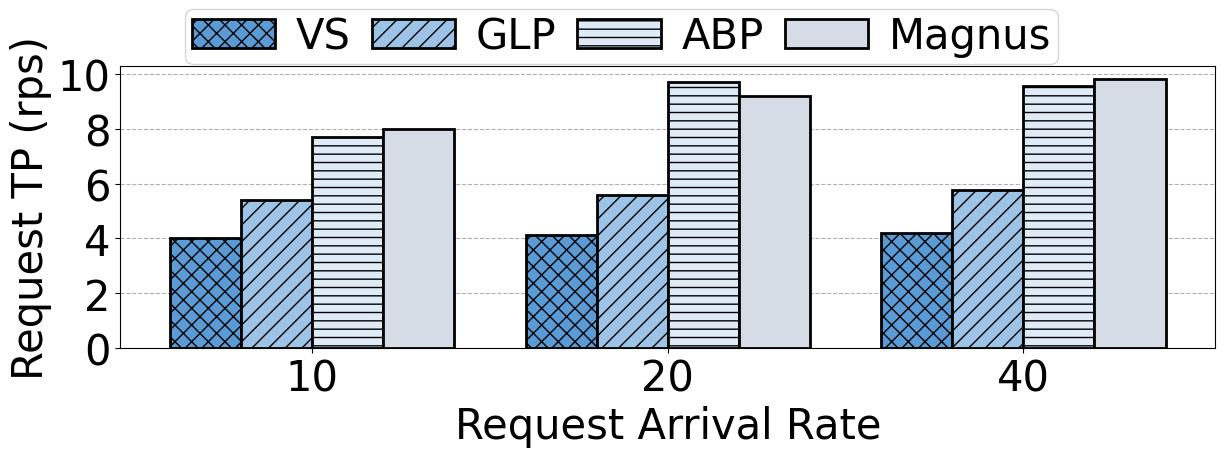}
		\caption{Request throughput.}
		\label{fig:abla_request_tp}
	\end{subfigure}
	\centering
	\begin{subfigure}{0.33\linewidth}
		\centering
		\includegraphics[width=1\linewidth]{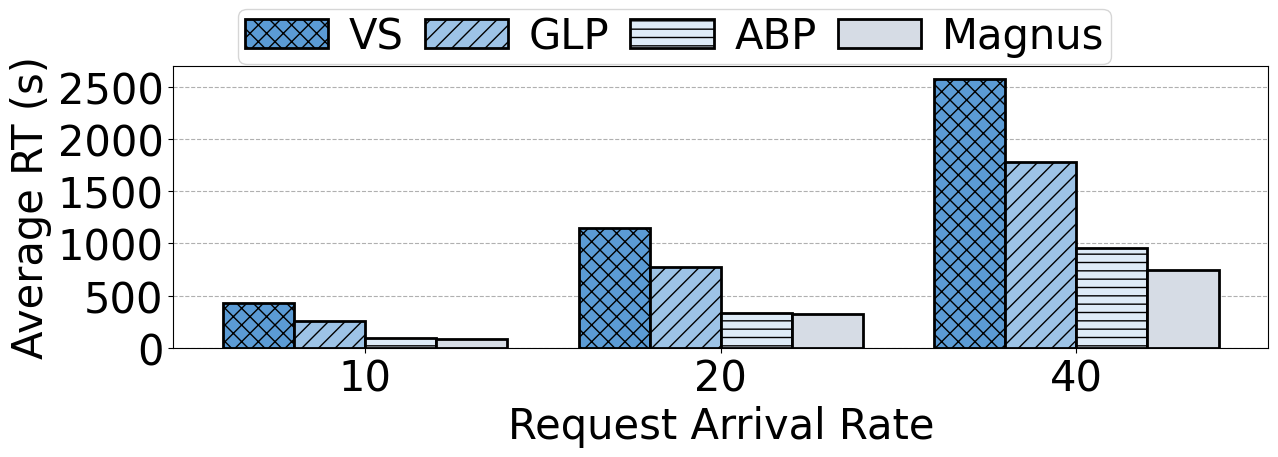}
		\caption{Average request response time.}
		\label{fig:abla_tail_res_time}
	\end{subfigure}
	\centering
	\begin{subfigure}{0.328\linewidth}
		\centering
		\includegraphics[width=1\linewidth]{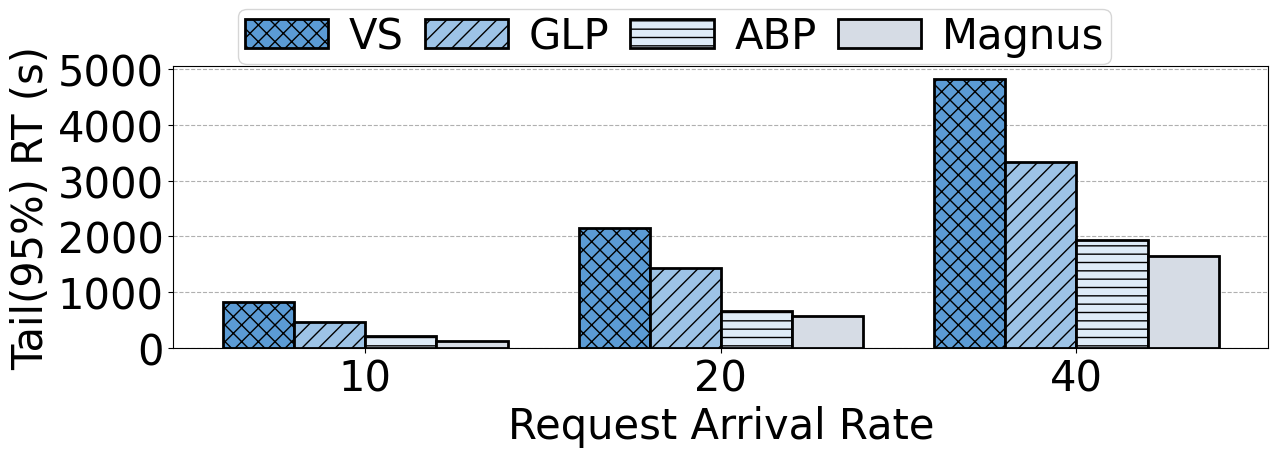}
		\caption{Tail (95\%) request response time.}
		\label{fig:abla_ave_res_time}
	\end{subfigure}
	\caption{Request-level performance benefits gained from each component of Magnus.}
	\label{fig:abla_request}
\end{figure*}

\subsection{Overall Performance}
In this subsection, we compare the performance of Magnus to the baselines under different request arrival rates. In the following experiments, seven ChatGLM-6B instances are separately deployed on seven NVIDIA 32GB V100 GPUs for serving requests. For Magnus, we set the WMA threshold $\Phi$ to 50,000 and the number of generation length predictors to 3 for parallel prediction. Therefore, 3 LaBSE models are deployed on the last V100 GPU. 

Fig. \ref{fig:eval_request} presents the performance of Magnus and the three baselines in terms of request throughput, average request response time, and tail response time under various request arrival rates. As shown in Fig. \ref{fig:eval_request}, Magnus always outperforms all the baselines, increasing the request throughput by 66\% to 234\%, and reducing the average and tail response time by 60.3\% to 89.7\% and 53.2\% to 91.7\% across various request arrival rates, respectively. In addition, as shown in Fig. \ref{fig:eval_token}, Magnus also outperforms the three baselines in terms of the token-level performance, increasing throughput of valid tokens and throughput of total tokens by 70\% to 240\% and 115\% to 489\%, respectively.

Magnus's superior performance in token throughput lies in its capability to predict the generation length of requests and leverage the predicted results to adaptively group requests into batches, which always enables a larger batch size than that of the baselines with a fixed batch size. Therefore, Magnus can take full advantage of the parallel computing capability of GPUs to achieve a higher token throughput.
Besides, since Magnus optimizes the wasted memory access (WMA) during batching, which effectively reduces the number of invalid tokens generated during inference, Magnus can achieve a high valid token throughput, thus achieving a high request throughput. 
Moreover, for Magnus, the throughput of total tokens, valid tokens, and requests continues to rise as the request arrival rate increases. This is because when the system is overloaded, more batches are waiting in the queue, and the newly arriving requests have more opportunities to join the queued batches, which further enlarges the batch size and improves the GPU utilization.

VSQ has the largest batch size among the three baselines, but its request throughput is always the smallest among these baselines, as shown in Fig. \ref{fig:eval_request}. There are two main reasons. First, quantization affects the generation quality of the LLM, and some unnecessary content is generated by VSQ. For example, when serving a batch with a request from the code translation (CT) application, the quantified LLM  may generate a lot of new codes after completing code translation, however, these new codes are redundant and increase the generation length of the request and the batch, thus increasing the batch serving time and decreasing the request throughput. Second, quantization brings additional computational overhead, which considerably declines the inference speed. Therefore, as shown in Fig. \ref{fig:eval_token}, although VSQ has a larger batch size than that of other baselines, its total token throughput is even smaller than VS, and the small token throughput further hurts the request throughput. Besides, although VSQ has a larger batch size compared with other baselines, the batch size is also fixed, failing to take full advantage of GPUs. Hence, Magnus outperforms VSQ in terms of request throughput by 193\% to 234\% under different request arrival rates.

CCB always has the smallest token throughput among all baselines as shown in Fig. \ref{fig:eval_token}, which is mainly because it introduces additional overhead in the inference process. For CCB, when a new request joins, the requests being served wait for the new request to complete the initialization phase, during the waiting, only one token of the newly arrived request is generated, which greatly hurts the token throughput. Since there is no invalid token generated by CCB, it is second only to Magnus in terms of valid token throughput as shown in Fig. \ref{fig:eval_token}. Thus its request throughput is also second only to Magnus's as shown in Fig. \ref{fig:eval_request}. Nevertheless, due to the small fixed batch size used by CCB, Magnus outperforms CCB in terms of request throughput by 66\% to 85.3\%.

For the response time, as presented in Fig. \ref{fig:eval_request}, Magnus exhibits a much shorter average and tail response time than all the baselines across various request arrival rates, not only because of the high request throughput yielded by the generation length prediction and adaptive batching but also because of the reduced average queuing time achieved by the serving time estimation-based HRRN batch scheduling.

VSQ processes more requests in parallel compared with other baselines, but its batch serving takes longer due to degradation in generation quality and additional computational overhead. Therefore, batches need to be queued for a long period of time, which results in the longest average and tail request response time among baselines as shown in Fig \ref{fig:eval_request}. Hence, compared with VSQ, Magnus reduces the average response time and tail response time by 80.4\% to 89.7\% and 76.9\% to 91.7\%  under different request rates, respectively.

CCB shows a shorter average request response time and tail response time than VS and VSQ  as shown in Fig. \ref{fig:eval_request}, because the completed request can be returned immediately and the queued requests can participate in the serving procedure in a timely manner, thus decreasing the queuing time of requests. However, compared with CCB, Magnus reduces the average response time and tail response time by 60.3\% to 73.5\% and 53.2\% to 77.5\% across various request arrival rates because of the high request throughput and low request queuing time.

\subsection{Ablation Studies}
We step-by-step validate the performance benefits obtained from each component of Magnus under different request rates. Firstly, we propose the \textit{\textbf{GLP}} strategy by adding the generation length predictor to VS. GLP leverages the WMA-directed batching to select the batch for requests to join, but it has a fixed batch size that limits the maximum number of requests in a batch to 7. Next, we lift the batch size limitation of GLP to achieve complete adaptive batching, and we call the new strategy \textit{\textbf{ABP}}. Finally, ABP is equipped with the serving time predictor and HRRN batch scheduler to make up Magnus.

Fig. \ref{fig:abla_request} presents the performance of VS, GLP, ABP, and Magnus in terms of request throughput, average response time, and tail response time under various request arrival rates. Besides, Fig. \ref{fig:abla_token} shows the token-level performance of the four strategies under various request arrival rates.

As shown in Fig. \ref{fig:abla_request}, under the same batch size limitation, GLP achieves a higher request throughput and shorter response time than VS because it can select the batch with the minimum WMA for requests based on the predicted generation lengths, thus effectively reducing generated invalid tokens and decreasing the batch serving time. This is also confirmed in Fig. \ref{fig:abla_token}, because of the reduced invalid tokens, GLP improves the valid token throughput by 35.7\% to 36.8\% compared with VS while they have almost the same total token throughput. 

By adaptively setting the batch size for each batch, ABP sets a large batch size for the batch with small length and generation length, thus making full use of the parallel computing capability of GPUs, and rapidly improves the token throughput by 106\% to 145\% on the basis of GLP as shown in Fig. \ref{fig:abla_token}, and hence it further improves the request throughput by 42.7\% to 74.3\%, and reduces the average response time and the tail response time by 46.4\% to 63.5\% and 42.2\% to 54.3\%, respectively, as shown in Fig. \ref{fig:abla_request}. 

Based on the serving time prediction-based HRRN scheduling, Magnus reduces the average queuing time of requests, and hence it reduces the average and tail response time by 5.1\% to 21.8\% and 13.7\% to 41.9\% compared with ABP as shown in Fig. \ref{fig:abla_request}, without affecting the request throughput.

\begin{figure}[t]
	\setlength{\belowcaptionskip}{-0.25cm} 
	\centering
	\begin{subfigure}{0.4925\linewidth}
		\centering
		\includegraphics[width=1\linewidth]{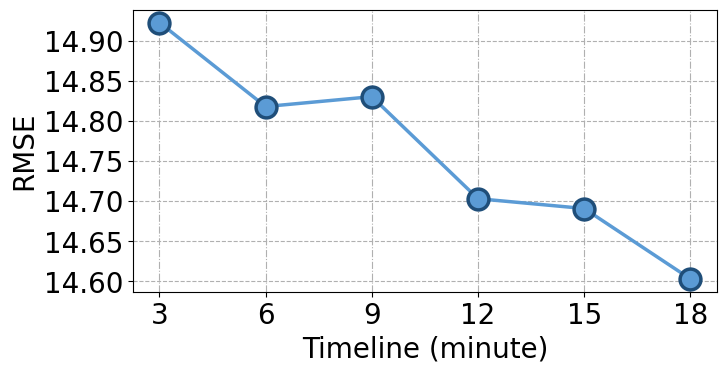}
		\caption{Generation length predictor.}
		\label{fig:acc_glp}
	\end{subfigure}
	\centering
	\begin{subfigure}{0.4925\linewidth}
		\centering
		\includegraphics[width=1\linewidth]{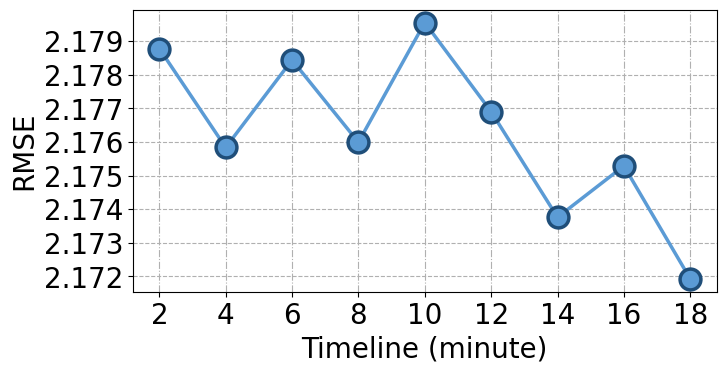}
		\caption{Serving time estimator.}
		\label{fig:acc_itp}
	\end{subfigure}
	\caption{Time-varying prediction error of predictors.}
	\label{fig:acc}
\end{figure}

Moreover, we also verify the effectiveness of continuous learning for the generation length predictor and serving time estimator Fig. \ref{fig:acc} depicts the root mean squared error (RMSE) of the generation length predictor and the serving time estimator over time.
As shown in Fig. \ref{fig:acc}, the accuracy of the two predictors is constantly refined by continuous learning.

\subsection{System Overhead}
In the experiments, on the latency side, the generation length prediction, batch packaging, serving time estimation, and batch scheduling take less than 0.03s, 0.001s, 0.001s, and 0.002s, respectively, which is negligible compared with the batch serving time that takes several seconds to hundreds of seconds. On the resource side, although the LaBSE model is deployed on an NVIDIA V100 32GB GPU in the experiments, since a LaBSE model only consumes less than 2GB GPU memory,  in the production environment, it can be individually deployed in the weak GPU with a small memory size, such as NVIDIA GTX1080 8GB GPU, without occupying the high-end GPU resources used for LLM deployment.

\section{Related Works}
\label{sec:related_work}

A variety of approaches have been proposed to improve the computing efficiency of LLM batch serving. We summarize these relevant studies by the following categories.

\textbf{Model compression}. 
Model weight quantization \cite{frantar2023optq, xiao2023smoothquant, shao2023omniquant, chee2024quip} converts the numerical weight precision of LLMs into fewer bits to reduce the memory footprint of LLM parameters. The technique of pruning \cite{xu2022dense, sun2023simple, liu2023deja} eliminates redundant components of LLMs to increase the inference speed. However, these methods can compromise the LLM generation quality and require either complete retraining or extra fine-tuning of LLMs, which is computationally expensive.

\textbf{Kernel optimization}. Kernel optimization \cite{aminabadi2022deepspeed, dao2022flashattention, kwon2023efficient, dao2023flashattention2, fu2022tcb, fang2021turbotransformers} accelerates LLM training and inference using highly efficient CUDA kernels, which enhances computational efficiency. These advanced low-level optimizations are orthogonal to Magnus, and they can enhance Magnus as well.

\textbf{Speculative sampling}. Some studies accelerate LLM inference with speculative sampling \cite{leviathan2023fast, xia2023speculative, sun2024spectr, kim2024speculative}, where a small model runs autoregressively to generate tokens and a large model occasionally refines the small model’s generated results in parallel. However, most speculative decoding algorithms do not support batch processing, which hurts the throughput.

\textbf{Request scheduling}. Existing serving systems \cite{triton, tensorflowserving} lack model-specific optimizations and serve requests in an FCFS manner with a fixed batch size which leads to severe computational inefficiency for LLM serving. Orca \cite{yu2022orca} and FastGen \cite{deepspeed-fastgen} utilize continuous batching to boost serving efficiency, but their conservative approaches to memory management act as a bottleneck for throughput. PiA \cite{zheng2024response} and $S^3$ \cite{jin2024s} propose to predict the request generation length to improve batch serving efficiency. However, they can only roughly predict the approximate range of request generation lengths, and cannot achieve token-level predictions.

\section{Conclusion}
\label{sec:conclusion}
In this paper, we propose Magnus to achieve efficient LLM serving for the LMaaS scenario, which predicts the request generation length based on the semantic features of the instruction and user input as well as the user input length. Magnus adaptively adjusts the batch size according to the predicted request generation lengths to make full use of the GPU's parallel computing capability, thus achieving a high request throughput. In addition, Magnus reduces the request response time via serving time estimation-based HRRN scheduling. Extensive experiments are conducted to demonstrate that Magnus can effectively reduce computational waste and fully utilize GPUs. In the future, we plan to extend Magnus to the scenario where LLM instances are deployed on heterogeneous GPUs. 

\section*{Acknowledgment}
We sincerely thank the anonymous reviewers for their insightful feedback and constructive suggestions. This work is supported in part by the Postgraduate Research \& Practice Innovation Program of Jiangsu Province (KYCX24\_0247) and the Ant Research Program of Ant Group.

\bibliographystyle{IEEEtran}
\bibliography{IEEEabrv,ref}

\end{document}